\documentclass[runningheads]{llncs}
\usepackage[T1]{fontenc}
\usepackage{graphicx}

\usepackage{hyperref}       
\usepackage{url}            
\usepackage{booktabs}       
\usepackage{amsfonts}       
\usepackage{nicefrac}       
\usepackage{microtype}      
\usepackage{xcolor}         
\usepackage[margin=1in]{geometry}

\usepackage{color}

\usepackage{amsmath}
\usepackage{amssymb}
\usepackage{mathtools}

\usepackage{latexsym,amsbsy,times}
\usepackage{bbm}
\usepackage{bm}
\usepackage{caption}
\usepackage{subcaption}
\usepackage{algorithm}
\usepackage{algpseudocode}
\usepackage{float}

\newcommand{\R}{\ensuremath{\mathbb{R}}}   

\newcommand{\E}{\ensuremath{\mathbb{E}}}

\newcommand{\floor}[1]{\left\lfloor #1 \right\rfloor}
\newcommand{\ceil}[1]{\left\lceil #1 \right\rceil}

\def \e {\varepsilon}

\def \cA {\mathcal{A}}

\def \cN {\mathcal{N}}

\def \cS {\mathcal{S}}

\DeclareMathOperator*{\argmin}{arg\,min}

\begin{document}
\title{TRUST: Transparent, Robust and Ultra-Sparse Trees}

\author{Albert Dorador}
\authorrunning{A. Dorador}
\institute{University of Wisconsin -- Madison,  Madison WI 53706, USA\\
\email{trustalgorithm.dev@gmail.com}\\
\url{https://adc-trust-ai.github.io/trust}}

\maketitle              
\begin{abstract}
Piecewise-constant regression trees remain popular for their interpretability, yet often lag behind black-box models like Random Forest in predictive accuracy. In this work, we introduce TRUST (Transparent, Robust, and Ultra-Sparse Trees), a novel regression tree model that combines the accuracy of Random Forests with the interpretability of shallow decision trees and sparse linear models. TRUST further enhances transparency by leveraging Large Language Models to generate tailored, user-friendly explanations. Extensive validation on synthetic and real-world benchmark datasets demonstrates that TRUST consistently outperforms other interpretable models -- including CART, Lasso, and Node Harvest -- in predictive accuracy, while matching the accuracy of Random Forest and offering substantial gains in both accuracy and interpretability over M5', a well-established model that is conceptually related.

\keywords{Interpretable Machine Learning \and Regression Trees \and Relaxed Lasso.}
\end{abstract}
\section{Introduction}
\label{sec:Intro}

Black-box models have been shown to be vulnerable to the `Clever Hans' phenomenon: they may predict the right answer (even occasionally out-of-sample) for the wrong reasons \cite{Ribeiro2016,Lapuschkin2019}, which makes them prone to failure in future cases.  Just like a non-exhaustive but otherwise arbitrarily large amount of examples does not prove a proposition, no amount of positive performance metrics can prove that a predictive machine learning algorithm is behaving as expected. Therefore, interpretability, in real-world, finite sample settings, might be the closest thing we have to a proof of performance. 

On the other hand, regression trees \cite{Morgan1963} excel at interpretability, particularly when they are short; in general, everything else being equal, the shorter they are, the easier they are to interpret \cite{Dorador2025}. A tree structure has a natural visualization, which tends to create good explanations in the sense of Miller \cite{Miller2019}: simple (as binary logic rules are  easy to understand),  contrastive (since one can always compare the different leaves and their corresponding root-to-leave path), and general (it provides explanations that apply to many events, by the very nature of binary splits).  If, in addition the tree is short, explanations are also selective, i.e. only a few reasons explain the result. 

Fitting a linear model in the nodes of the tree is one way to reduce the number of splits while typically improving accuracy.  The intuitive reason for the latter is that the hypothesis space of linear model trees includes piecewise-constant trees.

Linear model trees, being analogous to piecewise-linear functions, overcome one of the main drawbacks of traditional regression trees: their \textit{de facto} inability to (exactly) learn a linear function, as that would require an infinite amount of splits. For this reason, in the presence of a (perhaps piecewise) linear relation between inputs and outputs that is to be learned without excessive noise,  linear trees will typically produce considerably shorter trees for the same level of accuracy. 

After the development of the linear model tree \cite{Quinlan1992}, the present work proposes a further generalization of the regression tree model: TRUST (Transparent, Robust and Ultra-Sparse Trees), which fits a relaxed Lasso model \cite{Meinshausen2007} in the leaves, since linear regression is the special case of unpenalized relaxed Lasso. TRUST exploits synergies between regression trees and the relaxed Lasso, as  after having split on several covariates only a subset of all covariates may be relevant in the leaves,  which are then identified more efficiently than with stepwise selection (and with appropriate shrinkage).

The rest of this paper is structured as follows: Sections \ref{sec:T} through \ref{sec:US} discuss the transparency,  robustness and sparsity properties of TRUST, respectively.  Section \ref{sec:Res} presents our empirical results comparing the performance of TRUST against interpretable and black-box machine learning models,  on a mixture of benchmark and synthetic datasets of different characteristics. Section \ref{sec:Conclusion} concludes.

\section{TRUST is Transparent}
\label{sec:T}

\subsection{Global Transparency}

Global transparency (or interpretability) refers to the degree to which a human can understand the entire behavior of the model.  TRUST is equipped with two main tools to enhance global transparency: variable importance scores, as well as a tree diagram.

\subsubsection{Variable Importance Scores}

Let us begin by distinguishing between population feature importance and model feature importance \cite{Fisher2019}. The former refers to a feature’s true importance in the underlying population model, while the latter pertains to its importance within a specific model learned by an algorithm. These two notions may or may not align. Assessing population feature importance poses several challenges: by definition, the true population model is unknown, casting doubt on the reliability of any inferred importance scores. Philosophically, this notion is also problematic, as the (implied) uniqueness of the population model  cannot be guaranteed (Rashomon effect \cite{Breiman2001b}). In contrast, model feature importance offers practical value. Understanding the importance a model assigns to each feature is crucial for correcting biases, diagnosing errors, and fostering user trust. For these reasons, our focus is on model feature importance.

With this clear goal in mind, we delve into (model) variable importance scoring, which is a question that has gained considerable interest in the last two decades due to the success of black-box machine learning models and the corresponding desire to interpret them \cite{Loh2021bis}.  Despite the popularity of permutation methods \cite{Breiman2001},  they ignore

the correlation between covariates. This issue was the motivation behind the development of the so-called Ghost variable importance method \cite{Delicado2023}.  Another popular method called `leave-one-covariate-out' (LOCO), as its name suggests,  is based on feature occlusion, which makes it a computationally intensive method. 

However,  TRUST offers a fast method called `Ghost guide',  which, by combining ideas from the GUIDE \cite{Loh2021bis} and the Ghost variable importance methods,  provides uncertainty quantification,  negligible bias, and addresses the correlation ignorance of the permutation method.  

The first step is to calculate the Ghost variable importance scores for each feature as in Delicado and Peña \cite{Delicado2023}, with a modified score function. In order to do so, first we obtain the `ghost representation' of the $j$-th variable $v_j$, defined as $\hat v_j = \widehat{\E(v_j | X{_{-j}})}$, where $X{_{-j}}$ denotes the set of covariates excluding the $j$-th one. 

The original paper suggests using ordinary least squares (OLS) to estimate the previous expectation.  However, the choice of estimator should align with the model's hypothesis space. In our case -- penalized linear trees of typically small depth -- OLS is appropriate, as each variable's influence can plausibly be approximated by a linear combination of the others (to a large extent, at least\footnote{Any bias this choice might produce would be eliminated by our debiasing step.}). We compute the importance score for the $j$-th variable as the ratio of the model’s mean squared error (MSE) when the original $j$-th feature is replaced by its best linear reconstruction (a `ghost' feature), to the original MSE. This score reflects how much model accuracy deteriorates when that variable is substituted by its linear proxy. Our scoring function differs from that of the original authors, who use the ratio of the squared prediction difference to the original MSE -- an approach that may not reflect true loss in predictive quality, as large differences in predictions do not necessarily imply worse performance. Our modification ensures the score more directly measures the impact on predictive accuracy.

As in standard linear regression, all else being equal, a variable that is more correlated with the remaining variables is deemed less important for the model, better reflecting the importance of the variable in the model conditional on the presence of the rest of variables. 
Thus, if the fitted tree has depth zero the relative ranking of the features in Ghost guide coincides with the one we would obtain from the standardized coefficients in a linear model. However, as soon as the tree depth increases past zero, the two measures of importance start to diverge and a specialized scoring method becomes necessary.

The second step is the score debiasing step, adapted from Loh and Zhou \cite{Loh2021bis}. As in the original method, the response is permuted. However,  this may not be enough if the response has low entropy. For that reason, we also add Gaussian noise with zero mean and variance that is inversely proportional to the sample variance in the response vector,  ensuring the new contaminated response has the same mean and a similar but somewhat higher variance than the original response (although asymptotically equal). 
Now, step one is repeated substituting the original response vector by the new contaminated one.  Then, the original scores are divided by the new scores. This way,  if the scoring system is biased towards certain features, the updated scores should correct for that (as the new scores using a contaminated response would still unduly favor the same features). 

The final step is uncertainty quantification, adapted from Loh and Zhou \cite{Loh2021bis} too,  employing a similar idea (with our same modification as in the debiasing step) to simulate the null hypothesis of independence of the response with respect to all covariates.

\begin{figure}[htbp!]
\centering
\includegraphics[width=0.75\linewidth]{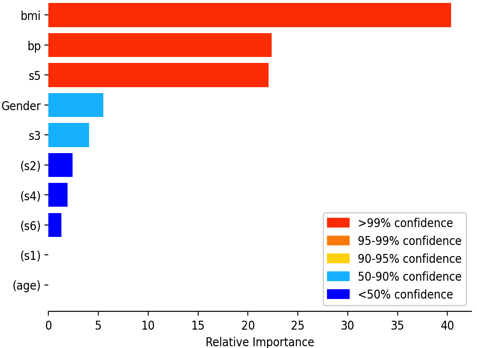}
\caption{Variable Importance Scores in the Diabetes dataset (type: `basic')}
\label{fig:Ghost-Diabetes}
\end{figure}

Then, the user is given a choice of three different plots (or none) to visualize the final importance scores along with their confidence level.  The default plot (Figure \ref{fig:Ghost-Diabetes}) offers a clean, more basic layout,  while the other two plot options also encode information about whether the association between a given feature and the response is positive or negative overall via Kendall's $\tau$.

Lastly,  Accumulated Local Effects (ALE) plots are included to show how the predicted response changes as a given feature changes, while accounting for feature correlations.  Compared to variable importance plots, ALE plots provide a more nuanced representation of feature effects,  which are not constrained to be monotone anymore.

\subsubsection{Tree Diagram}

Visualizing the tree structure is essential for understanding the model’s global behavior. To ensure clarity, special care was taken in the design: categorical splits can display the complement set when it is smaller, and redundant elements already shown in parent nodes are omitted to reduce visual clutter.

Figure \ref{fig:Tree-Boston} below shows an example tree diagram for the famous Boston dataset.

\begin{figure}[htbp!]
\centering
\includegraphics[width=0.67\linewidth]{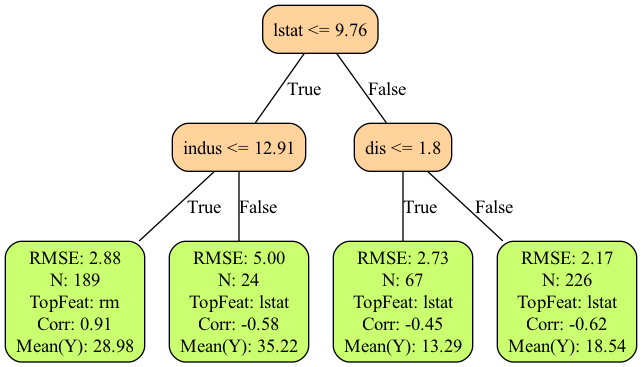}
\caption{Tree topology learnt in the Boston Housing dataset}
\label{fig:Tree-Boston}
\end{figure}

\subsection{Local Transparency}

Local transparency (or interpretability) refers to the degree to which a human can understand any given individual prediction made by the model. 

\begin{figure}[htbp!]
\centering
\includegraphics[width=0.8\linewidth]{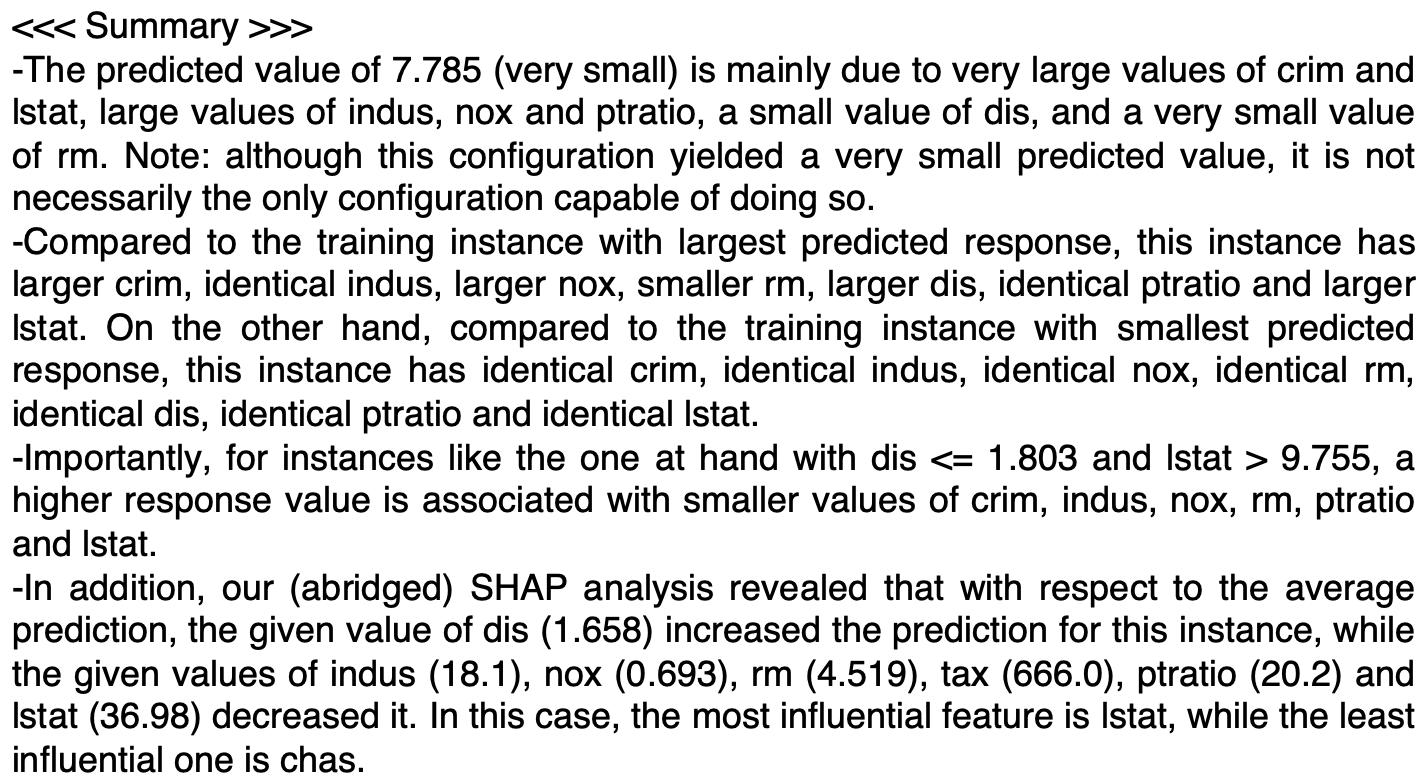}
\caption{Explanation summary for the prediction of an observation in the Boston dataset}
\label{fig:Summary-Boston}
\end{figure}

TRUST generates a report that includes the feature values of the observations with highest and lowest predicted values. This allows the user to compare the attributes of the requested instance with those of the instances with largest and smallest predictions (second bullet point in Figure \ref{fig:Summary-Boston}).
This type of explanation by comparison with a reference instance (`contrastive explanation') has been proposed since Lipton \cite{Lipton1990bis}, arguing that humans may be more interested in why a decision was made \textit{instead of another one},  rather than just knowing why the decision itself was made.

Then, the penalized linear model coefficients estimates for the leaf where the observation of interest lies are shown in a table,  along with approximate p-values. 

Next, TRUST carries out a SHAP \cite{Lundberg2017bis} analysis.  SHAP (Shapley additive explanations) is a model-agnostic local explanation tool that assigns each feature a positive or negative contribution for a particular prediction. Within the class of additive feature attribution methods, only methods based on Shapley values like SHAP possess the following desirable three properties: local accuracy, missingness and consistency.

At the end of its explanation report, TRUST outputs an automatically-generated text summary,  which includes the main highlights of the explanation (see Figure \ref{fig:Summary-Boston}).

\begin{figure}[H]
\centering
\includegraphics[width=0.67\linewidth]{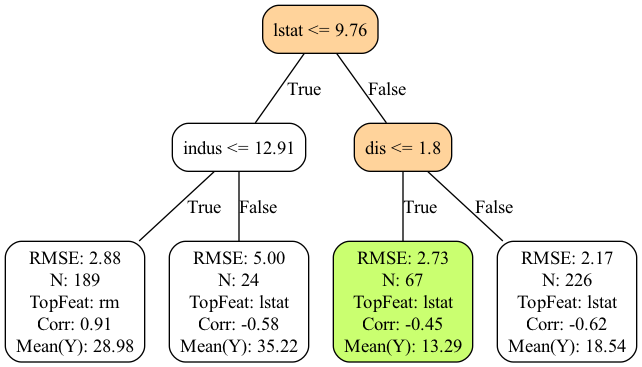}
\caption{Root-to-leaf plot for the first observation in the Boston dataset}
\label{fig:Root_to_leaf-Boston}
\end{figure}

In addition, a `root-to-leaf path' plot (Figure \ref{fig:Root_to_leaf-Boston}) is optionally displayed and saved.

The latest addition to TRUST's local transparency toolkit has been the successful integration of Google's Large Language Model (LLM) Gemini within the \texttt{explain} method. The integration of LLMs into supervised learning remains in its early stages, and this work is the first to leverage LLMs for generating custom prediction explanations.

How it works: the \texttt{explain} method gets the relevant model parameter values (splits and linear model coefficients in the relevant leaf) and communicates that information to the LLM, along with some prompts including explicit instructions to prevent the LLM from providing a solution that violates the split conditions, as that would push the observation to a different leaf with different coefficients. 

Assigning a persona to an LLM in one's prompt can be a useful technique to guide its response style, tone, and focus, potentially improving task performance, especially in creative writing and reasoning, but also in accuracy-based tasks \cite{Kong2024bis}, although care must be taken \cite{Zheng2024bis}. Allowing the user to choose a specific persona besides the default one (a `linear model tree expert') gives the user control on the style and complexity level of the responses, further tailoring the LLM's explanations to the user's needs.

Communication with the LLM is interactive, allowing users to ask multiple questions and get respective answers in a sequential fashion, in a fluid dialog with the LLM, which can ask the user clarifying questions to provide more relevant insights and actionable advice. The connection with the LLM is made via the  \href{https://github.com/google-gemini/generative-ai-python}{Gemini API}  for Python.

\begin{figure}[htbp!]
\centering
\includegraphics[width=\linewidth]{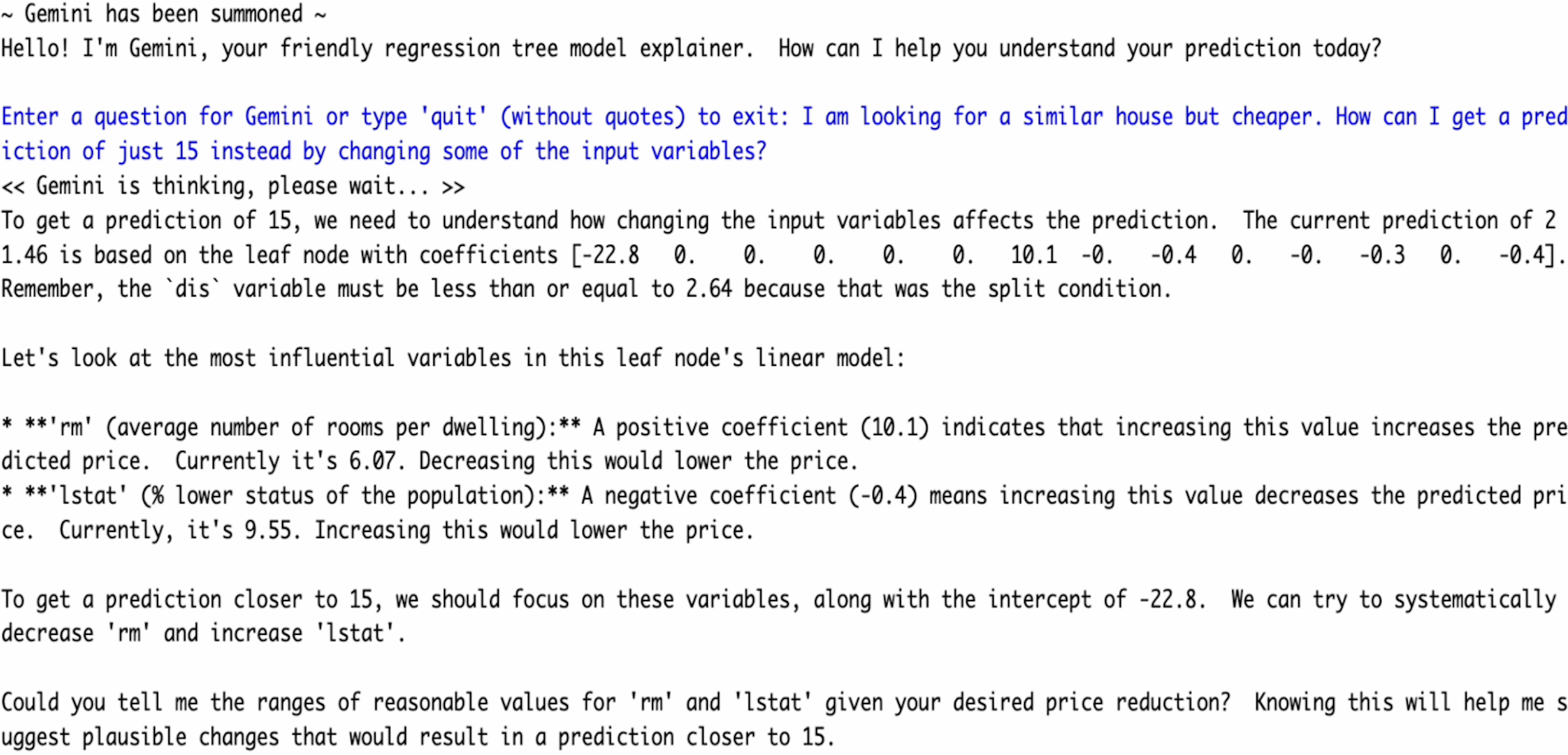}
\caption{Demonstration of Gemini's integration within TRUST in the Boston dataset}
\label{fig:Still_3}
\end{figure}

Below (Figure \ref{fig:Still_3})  is a still image from a video demonstration that is accessible on the  \href{https://adc-trust-ai.github.io/trust/}{author's website}.
The demonstration uses the well-known Boston Housing dataset that we have been using all along.

Figure \ref{fig:Still_3}  shows that the reasoning capabilities displayed by Gemini 1.5 Flash are very good even in non-trivial tasks like the one at hand.  Gemini displayed good judgment, when asking the user about acceptable ranges for `rm' and `lstat',  as this information can lead to more relevant LLM-generated solutions in fewer human-machine interactions. Lastly, Gemini automatically deduced by the name of the covariates that the dataset at hand was the famous Boston Housing dataset and hence was able to autonomously provide useful context regarding the relevant variables `rm' and `lstat'.

\section{TRUST is Robust}
\label{sec:R}

Model robustness, just like model transparency,  is often considered a requirement for machine learning trustworthiness \cite{Li2023}. Robustness refers to a model's ability to maintain a stable performance across varied conditions, including unexpected inputs \cite{Grote2023}. One of the main factors that challenges the robustness of a model is data bias, which arises when the assumption that the training data and the unseen data come from the same distribution (known as `closed-world' assumption) is violated. In real-world applications, there is often some degree of data bias, as obtaining training datasets that provide a complete and truthful picture of the data-generating process can be challenging \cite{Sugiyama2012}.

In this section, we will discover how TRUST tackles robustness from multiple angles.

\subsection{Out-of-Distribution Detection}
\label{sec:OOD}

As previously noted,  machine learning models are commonly trained under the close-world assumption, where test data is assumed to be drawn i.i.d. from the same distribution as the training data, referred to as `in-distribution'. However, when models are deployed in an open-world scenario test samples can, in contrast, exhibit a covariate shift which makes them `out-of-distribution' (OOD) \cite{Liu2020bis}, and thus should be handled with caution.  

To that end, TRUST provides the option of performing OOD detection at prediction time, warning the user about specific instances that appear to be OOD.  In line with the prevalent practice \cite{Zimmerer2022bis},  in TRUST predictions are generated regardless of their in- or out- of distribution assessment (perhaps with an adjustment, see next subsection). However, the accompanying OOD assessment empowers users to discard specific predictions or investigate potential data issues that may have triggered the OOD flag. 

There is a plethora of OOD detection methods,  whose relative strengths depend on the context and goals \cite{Yang2024}.  To identify OOD samples, TRUST uses a distanced-based approach: a modified Mahalanobis distance replacing the sample mean by the median to better reflect the goal of identifying test instances that are `far' from the training data, while accounting for covariate correlations that inform what `far' means. This substitution enhances robustness to outliers and reduces reliance on the assumption that the data-generating process is fully characterized by its first two moments (as in e.g. the Gaussian case).

In addition, a complementary analysis is performed, warning the user in case the given test sample exceeds either the minimum or maximum in-sample feature values,  reporting the precise percentage of said breach. This secondary warning is tightly related to extrapolation, which we discuss next. 

\subsection{Extrapolation Handling}
\label{sec:EE}

The notion of model `safety' has two main components: the expectation and the variance of the model's accuracy.  In this subsection we discuss a surprisingly simple solution to improve safety on both accounts while still allowing the model to extrapolate as needed.

In the context of regression trees, extrapolation occurs whenever the given instance we are requesting a prediction for lies within a leaf but outside the convex hull of the training data in that leaf. For example, one of the sets in a partition may include all values of a certain feature that are above $a\in \R$, but the largest such value in the training data was $b$, while the instance at hand has value $c > b$. This problem is discussed at length in Loh et al. \cite{Loh2007}, where the authors propose truncating the model predictions as a way to reduce extrapolation errors,  in expectation.  We generalize this truncation idea in three ways: first, by introducing a constant $t \in \R$, instead of using a fixed value; second, by using an asymetric measure of in-training response dispersion $S,S'$; lastly, by considering each leaf individually. Our modified truncation algorithm is then as follows:
\begin{equation}
\label{truncAlg}
\hat y = \max \left[\min\left(\hat f(x), \max(y_{\ell}) + tS(y_{\ell}) \right),  \min(y_{\ell}) - tS'(y_{\ell})\right]
\end{equation}
where $\min(y), \max(y)$ denote, respectively, the smallest and largest values of the response observed in the training set. We chose $S' = SD(y | y<median(y))$ and $S = SD(y | y \geq median(y))$, where $SD$ denotes the sample standard deviation. Then, $t$ is determined in-sample for efficiency reasons, and can take on large positive values as well as negative values, the former implying no truncation, while the latter implies a truncation above the minimum and below the maximum response in the training set.  Observe that, although the primary motivation of this framework was to get protection against extrapolation,  $t <0$ may yield interpolation protection as well.

\begin{figure}[htbp!]
\centering
\includegraphics[width=\linewidth]{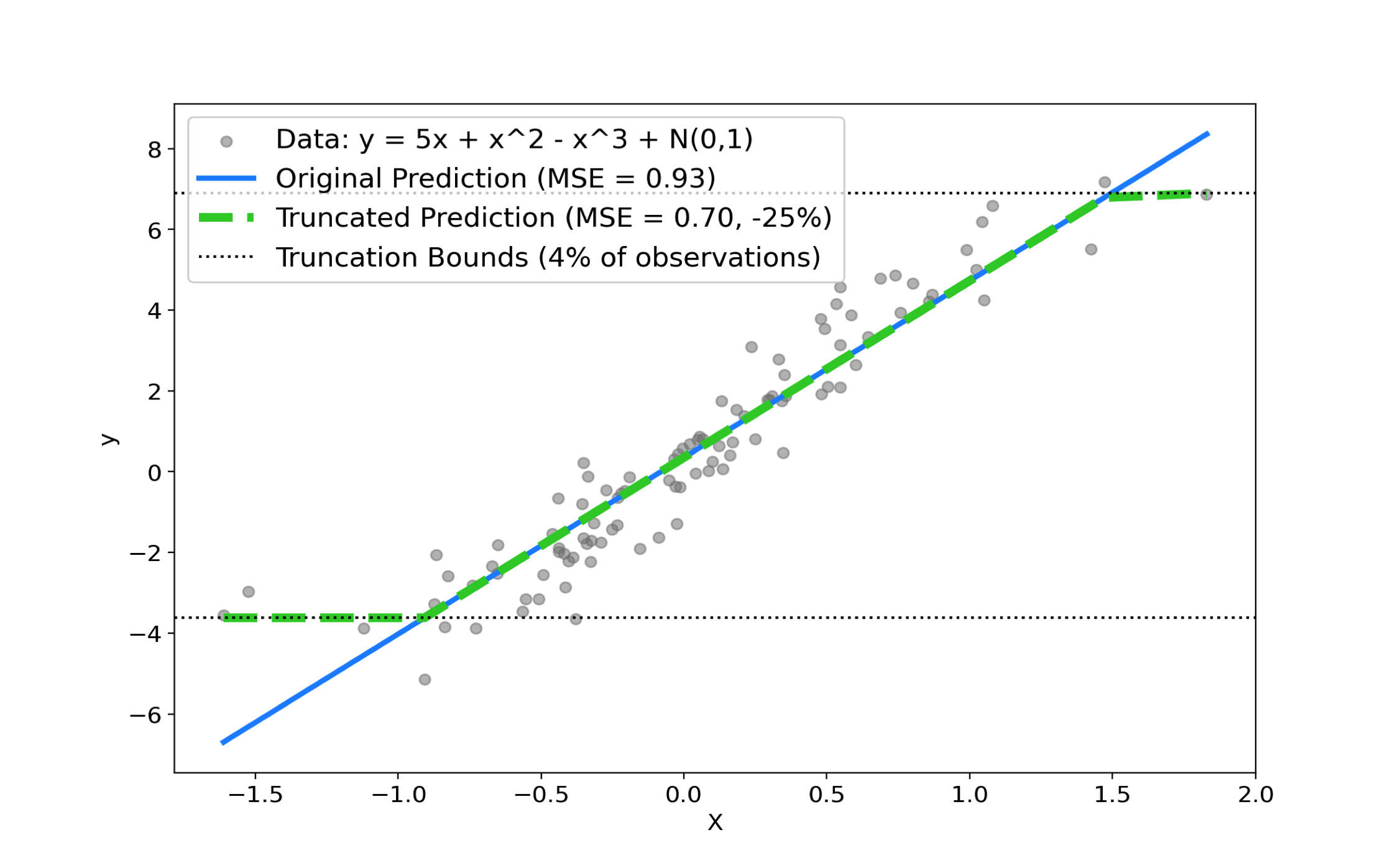}
\caption{An example of the potential benefits of truncating a linear model prediction}
\label{fig:Trunc_pred}
\end{figure}

As illustrated in Figure \ref{fig:Trunc_pred}, prediction truncation -- which modifies the original linear model fitted in the leaves, becoming piecewise linear -- can be very helpful.

\subsection{Missing Value Handling}
\label{sec:MV}

Working with real-world data often times requires handling missing values. There are several options: (a) excluding the instances with at least one missing value, (b) excluding the features with at least one missing value, (c) imputing the missing values, or (d) keeping the missing values.

The first two options are very simple but by definition waste the non-missing data in the deleted row or column, respectively. The impact of this drawback may be acceptable if the relative amount of discarded instances and / or features is small, which might be subjective and generally depends on a case-by-case basis. In addition, it ignores the potential information encoded in the missingness of the values itself.

Option (c) comprises a multitude of imputation methods, ranging from very simple ones like mean or median imputation, to more sophisticated ones like MICE \cite{vanBuuren2011} or AMELIA \cite{Honaker2011}.  Although these methods retain all non-missing data, they can be computationally intensive and/or yield low-quality imputations, as the imputation task can rival or exceed the complexity of the primary learning problem \cite{Loh2020b}.

Lastly, option (d) keeps the information contained in both the non-missing values as well as the missing values. The problem is that the vast majority of regression algorithms require the missing values to be imputed first.

In TRUST, option (d) is employed during data partitioning, while option (c) is used to fit the linear model in the nodes.  In other words, TRUST, just like GUIDE \cite{Loh2020}, can split on missing values, in one of three ways:
\begin{enumerate}
\item Split instances depending on whether or not they miss information about a given feature, disregarding any other criterion
\item Split instances depending on whether they do not miss information about a given feature and in that case the feature is below a certain threshold
\item Split instances depending on whether they miss information about a given feature or the feature is below a certain threshold
\end{enumerate}

Being able to split on missing values allows TRUST to learn any potential information encoded in the fact that certain values of given features are missing (in the common case that they are not missing completely at random), as well as warn the user about possible experiment design or data integrity issues. 
Then, global median imputation is used in the leaves to obtain a final linear model prediction.

\subsection{High-Dimensional Prediction}

Given a matrix of covariates $X\in \R^{n \times p}$,  the relaxed Lasso objective function in eq. \ref{eq:rL} is not strictly convex when $rank(X) < p$,  and so there can be multiple minimizers, which is not desirable, both from accuracy and interpretability perspectives. 
This situation immediately arises when the number of variables $p$ exceeds the number of observations $n$, which can easily happen in the leaves of regression trees.

The first measure we put in place is to ensure that the terminal nodes in TRUST have at least $p+2$ samples. Therefore, the only place where we might have $n \leq p+1$ is at the root of the tree. In this case, we need to decide how to handle the situation. If $n = p+1$, then Lasso, relaxed Lasso or even OLS will find the unique solution, but there is still substantial risk of overfitting, although less so with (relaxed) Lasso. If $n < p + 1$, then, although OLS can still be used (just not in the `conventional' way as the Gram matrix is not invertible),  we can always find a variable whose estimated coefficient is positive at one solution and negative at another. Although the sign inconsistency issue cannot happen in Lasso, the non-uniqueness of the solution might still be challenging.

For those reasons, if $n \leq p+1$ in the root of the tree, TRUST fits an elastic net model, which is always strictly convex in $\beta$, avoiding the potential issues described above with Lasso, and, especially, with OLS. Besides this advantage, the elastic net  overcomes the so-called saturation problem, which refers to the fact that when the Lasso solution is unique it can never assign more than $n$ non-zero coefficients, which could be problematic, particularly when $p$ is significantly larger than $n$. 

Note that the traditional elastic net indeed belongs to the family of models that TRUST naturally fits: depth-$d$ trees (where $d\geq 0$) with linear models in the leaves featuring a generalized form of elastic net where Lasso is replaced by relaxed Lasso.  In this setting, fitting a traditional elastic net corresponds to fitting a depth-0 TRUST tree with $\theta = 1$ in the relaxed Lasso objective function (eq. \ref{eq:rL}) and nonzero $\ell_2$ penalty.

\section{TRUST is Ultra-Sparse}
\label{sec:US}

By fitting a relaxed Lasso model \cite{Meinshausen2007bis}  in the leaves of the tree, we may achieve as sparse a linear model as possible without losing accuracy.

The motivation behind the development of relaxed Lasso was to separately control the two distinct (but related) effects of $\ell_1$ regularization, namely, model selection on the one hand and shrinkage estimation on the other hand. Unlike in the original Lasso, where both effects are controlled by a single parameter, in relaxed Lasso two parameters are used.  For example, in certain situations it might be optimal to  estimate the coefficients of all selected variables without any shrinkage at all.

The relaxed Lasso estimator is defined for $\lambda \geq 0$ and $\theta \in (0,1]$ as
\begin{equation}
\label{eq:rL}
\hat \beta_{\lambda, \theta} = \argmin_\beta \frac{1}{n} \sum_{i=1}^n \left(Y_i-X_i^T\beta \odot \mathbbm{1}_{\cS_\lambda} \right)^2 + \theta \lambda ||\beta||_1
\end{equation}
where $\mathbbm{1} \in \{0,1\}^p$ such that for all $k \in \{1,\ldots, p\}$,  $\mathbbm{1}_k = 1$ if and only if $k \in \cS_\lambda$ i.e. it is a $p$-dimensional binary vector that indicates whether the $k$-th covariate is active or not as a result of the choice of $\lambda$. Then, by using the Hadamard product, denoted by $\odot$,  the $k$-th coordinate of $\beta\odot \mathbbm{1}_{\cS_\lambda}$ is zero if the $k$-th covariate is not active, and $\beta_k$ otherwise.

As one can see, $\lambda$ controls variable selection while $\theta$ controls shrinkage. Observe that the original Lasso is a special case of the relaxed Lasso where $\theta = 1$. For $\theta < 1$,  the shrinkage in the selected model is reduced compared to ordinary Lasso estimation. Further, the case $\theta \to 0$ corresponds to a hybrid where Lasso is used for variable selection only and OLS is used to estimate the active coefficients. Lastly,  $\lambda = 0$ naturally corresponds to OLS, which again may be viewed as a special case of the (relaxed) Lasso. 

Theoretical and numerical results in Meinshausen \cite{Meinshausen2007bis} show that the relaxed Lasso tends to produce sparser models with equal or lower mean squared error than the ordinary Lasso estimator for high-dimensional data. 
Moreover, under regularity conditions, (i) relaxed Lasso enjoys faster convergence rates than ordinary Lasso (this is because noise variables can be prevented from entering the estimator with a high value of the penalty parameter $\lambda$,  while the coefficients of selected variables can be estimated at the usual $\sqrt{n}$ of OLS); (ii) choosing the parameters $\lambda, \theta$ by $k$-fold cross-validation is enough to guarantee that the relaxed Lasso is variable-selection consistent, unlike in Lasso.

\section{Empirical Results}
\label{sec:Res}

\subsection{Setup}
\label{sec:Final_sim_setup}

On the one hand, we have tested five synthetic datasets with four variations: smaller / larger sample (the latter indicated by the suffix `2' in tables),  smaller / larger noise (the latter denoted by the suffix `N'). Each variation is repeated with 2 different random seeds (123 and 321), across 10 cross-validation folds. 
\begin{enumerate}
\item Correlated: $y = X_1X_2^2 + X_2\exp(X_3) - X_3X_4^3 + \floor{X_1}\cos(X_4) + \varepsilon$, 

where $\left[X_1, X_2,X_3, X_4 \right] \sim \cN\left(
\begin{pmatrix}
0 \\
0 \\
0 \\
0
\end{pmatrix},  
\begin{pmatrix}
1 & -0.3 & 0.5 & 0.2\\
-0.3 & 1 & 0.6 & 0.5\\
0.5 & 0.6 & 1 & 0.8\\
0.2 & 0.5 & 0.8 & 1
\end{pmatrix}
\right)$

and $\varepsilon_i \overset{\mathrm{i.i.d.}}{\sim} \cN(0, 1 \text{ or }5^2)$ for all $i\in \{1,2,\ldots, 500 \text{ or } 5000\}$. Four noise variables  have been included as well.
\item Friedman: $y = 10\sin(\pi X_1X_2) + 20(X_3-0.5)^2 + 10X_4 + 5X_5 + \varepsilon$, where $X_j$ for all $j$ is uniformly distributed in $(0,1)$ and $\varepsilon_i \overset{\mathrm{i.i.d.}}{\sim} \cN(0, 1 \text{ or }5^2)$ for all $i\in \{1,2,\ldots, 500 \text{ or } 5000\}$. Five noise variables have been included.
\item Max: $y = 5\max(1 + X_1 + X_2, 0) + \varepsilon$, where $X_1,X_2 \sim \cN(0,1)$ and  $\varepsilon_i \overset{\mathrm{i.i.d.}}{\sim} \cN(0, 1 \text{ or }5^2)$ for all $i\in \{1,2,\ldots, 500 \text{ or } 5000\}$.  Includes two noise variables.
\item Sparse: $y = \sum_{j=1}^{50} \beta_j X_j + \varepsilon$ where $\beta_j = 0$ for all $j \in \{1,2,\ldots, 50\} \setminus \cA$,  $\cA = \{1,2,3,4,5\}$,  with $X_j \sim \cN(0,1)$ for all $j$, and $\varepsilon_i \overset{\mathrm{i.i.d.}}{\sim} \cN(0, 5 \text{ or }50^2)$ for all $i\in \{1,2,\ldots, 200 \text{ or } 2000\}$.  The much simpler functional form considered here justifies the comparatively higher noise levels tested.
\item Steps: $y = 10\ceil{\frac{X_1+X_2}{2}} + \varepsilon$, where $X_1,X_2 \sim \cN(0,1)$ and  $\varepsilon_i \overset{\mathrm{i.i.d.}}{\sim} \cN(0, 1 \text{ or }5^2)$ for all $i\in \{1,2,\ldots, 500 \text{ or } 5000\}$.  Two noise variables have been included.
\end{enumerate}

On the other hand, we have also tested 40 real-world datasets (Section \ref{sec:acc_res}),  each repeated with 2 different random seeds (123 and 321) across 10 cross-validation folds. 

TRUST and 5 other machine learning algorithms are evaluated with default parameters,  and, when applicable, one-hot encoding of categorical variables and median-imputation of missing values.  TRUST is by default limited to fitting at most 16 leaves. The alternative algorithms include 4 interpretable models and a black box: 
(i) M5'\cite{Wang1997}, an enhanced implementation of the linear model tree algorithm by Quinlan \cite{Quinlan1992}, R package RWeka, version 0.4-46; (ii) Lasso \cite{Tibshirani1996}, $\ell_1$-regularized linear regression, R package glmnet, version 4.1-8; (iii) NodeHarvest (NH) \cite{Meinshausen2010}, a sparse node selector from an initial random forest, R package nodeHarvest by the same author, version 0.7-3 (archived); (iv) CART \cite{Breiman1984}, piecewise-constant trees, R package rpart, version 4.1-19; (v) Random Forest (RF) \cite{Breiman2001},  an ensemble of 500 decision trees, R package randomForest,  version 4.7-1.2.

\subsection{Results}
\label{sec:acc_res}

First we show global accuracy metrics across all 60 datasets, and then we break them down by dataset type based on sample size, dimensionality and estimated noise level. The most accurate models, up to two decimal places, are shown in bold. Note: \textit{N/A} denotes that the measurement is not available due to algorithm failure.

\begin{table}[H]
\caption{Mean unexplained variance and mean ranks by model across 60 datasets}
\centering
\begin{tabular}{ l  c  c  c  c  c  c }
\toprule
Metric & TRUST & RF & M5' & CART & Lasso & NH \\ \midrule
Mean & \textbf{0.33} & 0.38 & 0.64 & 0.51 & 0.43 & 0.53 \\ 
Std & 0.03 & 0.03 & 0.04 & 0.03 & 0.04 & 0.03 \\ 
Mean Rank & \textbf{1.70} & 2.40 & 4.50 & 4.53 & 3.13 & 4.45 \\ 
Std Rank & 0.09 & 0.17 & 0.20 & 0.12 & 0.19 & 0.18 \\ 
\bottomrule
\end{tabular}
\label{tab:results}
\end{table}

Therefore, globally TRUST is at least as accurate as Random Forest (RF). Compared to other interpretable models, TRUST is more accurate, especially when compared to a similar model tree algorithm like M5', which also is significantly less interpretable, as it fitted a median of 109 coefficients per dataset, while TRUST's median is only 17.

\begin{table}[H]
\caption{Mean unexplained variance by model across 10 folds and 2 random seeds,  under a larger sample size (n), higher dimensionality (p) and higher estimated noise ($\hat{\sigma}_{\e}$)}
\centering
\begin{tabular}{ l l r  r  r  c  c  c  c  c  c  c }
\toprule
Dataset & Source & n & p & $\hat{\sigma}_{\e}$ & Missing & TRUST & RF & M5' & CART & Lasso & NH \\ \midrule
Crime & UCI repository & 1994 & 100 & 57.0 & 0 & 0.34 & 0.34 & 0.33 & 0.45 & \textbf{0.32} & 0.42 \\ 
Doctor & R package Ecdat & 5190 & 14 & 235.9 & 0 & \textbf{0.77} & 0.79 & 0.81 & 0.82 & 0.78 & \textbf{0.77} \\ 
FriedmanN2 & Synthetic & 5000 & 10 & 34.3 & 0 & \textbf{0.55} & 0.56 & 0.71 & 0.71 & 0.64 & 0.73 \\ 
Hours & R package Ecdat & 22272 & 11 & 56.4 & 0 & \textbf{0.60} & \textbf{0.60} & 0.61 & 0.66 & 0.64 & \textit{N/A} \\ 
Rent & R package Catdata & 2053 & 11 & 25.2 & 0 & \textbf{0.38} & 0.39 & 0.39 & 0.51 & \textbf{0.38} & \textit{N/A} \\ 
SparseN2 & Synthetic & 2000 & 50 & 1923.8 & 0 & \textbf{0.12} & 0.24 & 0.13 & 0.46 & \textbf{0.12} & 0.51 \\ 
\midrule
Mean & & 6418 & 33 & 388.8 & 0 & \textbf{0.46} & 0.49 & 0.50 & 0.60 & 0.48 & 0.61 \\ 
Std & & 3230 & 15 & 308.7 & 0 & 0.09 & 0.08 & 0.10 & 0.06 & 0.10 & 0.08 \\ 
Mean Rank & &  &  &  &  & \textbf{1.33} & 2.83 & 3.33 & 5.17 & 2.17 & 4.50 \\ 
Std Rank & &  &  &  &  & 0.33 & 0.48 & 0.42 & 0.31 & 0.54 & 1.19 \\ 
\bottomrule
\end{tabular}
\label{tab:resultsG1}
\end{table}

\begin{table}[H]
\caption{Mean unexplained variance by model across 10 folds and 2 random seeds,  under a larger sample size (n), higher dimensionality (p) and lower estimated noise ($\hat{\sigma}_{\e}$)}
\centering
\begin{tabular}{ l l r  r  r  c  c  c  c  c  c  c }
\toprule
Dataset & Source & n & p & $\hat{\sigma}_{\e}$ & Missing & TRUST & RF & M5' & CART & Lasso & NH \\ \midrule
Bikes & UCI repository & 17379 & 13 & 24.7 & 0 & 0.11 & \textbf{0.08} & 0.19 & 0.27 & 0.32 & \textit{N/A} \\ 
Computer & R package Ecdat &  6259 & 9 & 5.0 & 0 & 0.09 & \textbf{0.08} & >1 & 0.32 & 0.23 & \textit{N/A} \\ 
Debt & R package Ecdat & 2380 & 12 & 18.6 & 0 & \textbf{0.51} & 0.54 & 0.57 & 0.62 & 0.50 & 0.63 \\ 
Diamonds & R package ggplot2 & 53940 & 9 & 15.0 & 0 & 0.04 & \textbf{0.02} & 0.74 & 0.11 & 0.08 & \textit{N/A} \\ 
Friedman2 & Synthetic & 5000 & 10 & 7.0 & 0 & \textbf{0.06} & 0.11 & 0.36 & 0.40 & 0.28 & 0.46 \\ 
Parkinson & UCI repository & 5875 & 20 & 5.4 & 0 & \textbf{0.11} & 0.14 & 0.06 & 0.26 & 0.83 & 0.71 \\ 
School & R package Ecdat & 3010 & 25 & 5.8 & 1009 & \textbf{0.70} & \textbf{0.70} & 0.69 & 0.75 & \textbf{0.70} & 0.76 \\ 
Sparse2 & Synthetic & 2000 & 50 & 93.8 & 0 & \textbf{0.00} & 0.13 & 0.00 & 0.36 & \textbf{0.00} & 0.44 \\ 
Vietnam & The World Bank & 27765 & 11 & 15.1 & 0 & 0.54 & \textbf{0.50} & 0.63 & 0.62 & 0.81 & 0.68 \\ 
Wine & UCI repository & 6497 & 11 & 9.0 & 0 & 0.66 & \textbf{0.46} & >1 & 0.76 & 0.71 & 0.75 \\ 
\midrule
Mean & & 13011 & 17 & 19.9 & 101 & \textbf{0.28} & \textbf{0.28} & 0.52 & 0.45 & 0.45 & 0.63 \\ 
Std & & 5217 & 4 & 8.5 & 101 & 0.09 & 0.08 & 0.11 & 0.07 & 0.10 & 0.05 \\ 
Mean Rank & &  &  &  &  & \textbf{1.80} & 1.90 & 3.40 & 4.40 & 3.30 & 5.43 \\ 
Std Rank & &  &  &  &  & 0.13 & 0.35 & 0.58 & 0.22 & 0.58 & 0.30 \\ 
\bottomrule
\end{tabular}
\label{tab:resultsG2}
\end{table}

\begin{table}[H]
\caption{Mean unexplained variance by model across 10 folds and 2 random seeds,  under a larger sample size (n), lower dimensionality (p) and higher estimated noise ($\hat{\sigma}_{\e}$)}
\centering
\begin{tabular}{ l l r  r  r  c  c  c  c  c  c  c }
\toprule
Dataset & Source & n & p & $\hat{\sigma}_{\e}$ & Missing & TRUST & RF & M5' & CART & Lasso & NH \\ \midrule
Cars & Kaggle & 4340 & 6 & 73.9 & 0 & 0.44 & \textbf{0.34} & >1 & \textit{N/A} & 0.54 & \textit{N/A} \\  
CorrelatedN2 & Synthetic & 5000 & 8 & 305.4 & 0 & \textbf{0.42} & 0.45 & >1 & 0.56 & 0.92 & 0.59 \\ 
Food & R package Ecdat & 23971 & 5 & 33.1 & 0 & \textbf{0.58} & 0.65 & 0.64 & 0.68 & 0.65 & 0.69 \\ 
Income & R package carData & 4147 & 4 & 41.3 & 189 & \textbf{0.64} & 0.69 & 0.69 & 0.68 & 0.71 & 0.70 \\ 
MaxN2 & Synthetic & 5000 & 4 & 85.2 & 0 & \textbf{0.44} & 0.47 & 0.57 & 0.54 & 0.49 & 0.58 \\ 
StepsN2 & Synthetic & 5000 & 4 & 92.7 & 0 & 0.41 & \textbf{0.38} & 0.56 & 0.47 & 0.41 & 0.54 \\
\midrule
Mean & & 7910 & 5 & 105.3 & 32 & \textbf{0.49} & 0.50 & 0.74 & 0.59 & 0.62 & 0.62 \\ 
Std & & 3216 & 1 & 41.2 & 32 & 0.04 & 0.06 & 0.08 & 0.04 & 0.07 & 0.03 \\ 
Mean Rank & &  &  &  &  & \textbf{1.33} & 1.83 & 4.83 & 3.80 & 3.33 & 5.00 \\ 
Std Rank & &  &  &  &  & 0.21 & 0.31 & 0.65 & 0.37 & 0.42 & 0.45 \\ 
\bottomrule
\end{tabular}
\label{tab:resultsG3}
\end{table}

\begin{table}[H]
\caption{Mean unexplained variance by model across 10 folds and 2 random seeds,  under a larger sample size (n), lower dimensionality (p) and lower estimated noise ($\hat{\sigma}_{\e}$)}
\centering
\begin{tabular}{ l l r  r  r  c  c  c  c  c  c  c }
\toprule
Dataset & Source & n & p & $\hat{\sigma}_{\e}$ & Missing & TRUST & RF & M5' & CART & Lasso & NH \\ \midrule
Abalone & UCI repository & 4177 & 8 & 18.5 & 0 & 0.45 & \textbf{0.44} & 0.95 & 0.56 & 0.47 & 0.61 \\ 
California & StatLib & 20640 & 8 & 18.1 & 0 & 0.24 & \textbf{0.18} & >1 & 0.47 & 0.40 & 0.51 \\ 
Correlated2 & Synthetic & 5000 & 8 & 43.3 & 0 & \textbf{0.12} & 0.17 & >1 & 0.33 & 0.89 & 0.43 \\ 
London & R package bsamGP & 5113 & 6 & 10.1 & 0 & 0.41 & \textbf{0.36} & >1 & 0.46 & 0.56 & \textit{N/A} \\ 
Max2 & Synthetic & 5000 & 4 & 17.1 & 0 & \textbf{0.03} & 0.06 & 0.45 & 0.15 & 0.12 & 0.28 \\ 
Satisfaction & Eurostat & 5609 & 4 & 11.3 & 0 & \textbf{0.59} & 0.67 & 0.90 & 0.66 & 0.59 & 0.81 \\ 
Steps2 & Synthetic & 5000 & 4 & 38.8 & 0 & 0.16 & \textbf{0.09} & 0.57 & 0.22 & 0.16 & 0.36 \\ 
Wage & R package ISLR & 3000 & 8 & 5.9 & 0 & \textbf{0.62} & 0.63 & 0.64 & 0.71 & 0.64 & 0.72 \\ 
\midrule
Mean & & 6692 & 6 & 20.4 & 0 & \textbf{0.33} & \textbf{0.33} & 0.81 & 0.45 & 0.48 & 0.53 \\ 
Std & & 2012 & 1 & 4.8 & 0 & 0.08 & 0.08 & 0.08 & 0.07 & 0.09 & 0.07 \\ 
Mean Rank & &  &  &  &  & \textbf{1.50} & 1.75 & 5.50 & 3.75 & 3.00 & 5.00 \\ 
Std Rank & &  &  &  &  & 0.19 & 0.37 & 0.38 & 0.25 & 0.42 & 0.22 \\ 
\bottomrule
\end{tabular}
\label{tab:resultsG4}
\end{table}

\begin{table}[H]
\caption{Mean unexplained variance by model across 10 folds and 2 random seeds,  under a smaller sample size (n), higher dimensionality (p) and higher estimated noise ($\hat{\sigma}_{\e}$)}
\centering
\begin{tabular}{ l l r  r  r  c  c  c  c  c  c  c }
\toprule
Dataset & Source & n & p & $\hat{\sigma}_{\e}$ & Missing & TRUST & RF & M5' & CART & Lasso & NH \\ \midrule
Baseball & StatLib & 263 & 19 & 47.4 & 0 & 0.56 & \textbf{0.40} & 0.71 & 0.62 & 0.63 & 0.50 \\ 
College & R package ISLR & 775 & 17 & 34.9 & 0 & \textbf{0.10} & 0.11 & >1 & 0.18 & 0.09 & \textit{N/A} \\ 
Diabetes & UCI repository & 442 & 10 & 35.3 & 0 & \textbf{0.52} & 0.56 & 0.60 & 0.69 & 0.53 & 0.6 \\ 
FriedmanN & Synthetic & 500 & 10 & 36.1 & 0 & 0.65 & \textbf{0.64} & 0.73 & 0.76 & 0.65 & 0.74 \\ 
Ozone & UCI repository & 330 & 9 & 29.8 & 0 & 0.28 & \textbf{0.26} & >1 & 0.38 & 0.33 & 0.32 \\ 
SparseN & Synthetic & 200 & 50 & 423.2 & 0 & \textbf{0.16} & 0.47 & 0.19 & 0.64 & 0.17 & 0.56 \\ 
\midrule
Mean & & 418 & 19 & 101.1 & 0 & \textbf{0.38} & 0.41 & 0.71 & 0.55 & 0.40 & 0.54 \\ 
Std & & 84 & 6 & 64.5 & 0 & 0.09 & 0.08 & 0.12 & 0.09 & 0.10 & 0.07 \\ 
Mean Rank & &  &  &  &  & \textbf{2.00} & 2.33 & 5.00 & 4.67 & 2.67 & 3.00 \\ 
Std Rank & &  &  &  &  & 0.58 & 0.67 & 0.58 & 0.67 & 1.20 & 1.00 \\ 
\bottomrule
\end{tabular}
\label{tab:resultsG5}
\end{table}

\begin{table}[H]
\caption{Mean unexplained variance by model across 10 folds and 2 random seeds,  under a smaller sample size (n), higher dimensionality (p) and lower estimated noise ($\hat{\sigma}_{\e}$)}
\centering
\fontsize{10pt}{10pt}\selectfont
\begin{tabular}{ l l r  r  r  c  c  c  c  c  c  c }
\toprule
Dataset & Source & n & p & $\hat{\sigma}_{\e}$ & Missing & TRUST & RF & M5' & CART & Lasso & NH \\ \midrule
BMI & R package alr4 & 136 & 11 & 0.1 & 0 & \textbf{0.01} & 0.40 & >1 & 0.66 & \textbf{0.01} & 0.55 \\ 
Boston & R package MASS & 506 & 13 & 3.9 & 0 & 0.22 & \textbf{0.12} & >1 & 0.28 & 0.29 & 0.24 \\ 
Friedman & Synthetic & 500 & 10 & 6.8 & 0 & \textbf{0.16} & 0.22 & 0.32 & 0.40 & 0.28 & 0.45 \\ 
Riboflavin & R package hdi & 71 & 4088 & 4.7 & 0 & 0.33 & 0.57 & 0.97 & >1 & \textbf{0.32} & 0.62 \\ 
Riboflavin2 & R package hdi  & 71 & 100 & 7.1 & 0 & 0.33 & 0.38 & 0.52 & >1 & \textbf{0.31} & 0.50 \\ 
Sparse & Synthetic & 200 & 50 & 93.8 & 0 & \textbf{0.00} & 0.35 & 0.03 & 0.53 & \textbf{0.00} & 0.47 \\ 
Student & UCI repository & 649 & 30 & 20.2 & 0 & \textbf{0.75} & 0.69 & 0.82 & 0.82 & \textbf{0.75} & 0.78 \\ 
Student2 & UCI repository & 649 & 32 & 9.5 & 0 & 0.16 & 0.16 & \textbf{0.15} & 0.19 & 0.16 & 0.22 \\ 
\midrule
Mean & & 348 & 542 & 18.3 & 0 & \textbf{0.25} & 0.36 & 0.60 & 0.61 & 0.27 & 0.48 \\ 
Std & & 90 & 507 & 11.0 & 0 & 0.08 & 0.07 & 0.14 & 0.11 & 0.08 & 0.07 \\ 
Mean Rank & &  &  &  &  & \textbf{1.63} & 2.38 & 4.38 & 5.25 & 2.00 & 4.50 \\ 
Std Rank & &  &  &  &  & 0.18 & 0.38 & 0.60 & 0.25 & 0.50 & 0.38 \\ 
\bottomrule
\end{tabular}
\label{tab:resultsG6}
\end{table}

\begin{table}[H]
\caption{Mean unexplained variance by model across 10 folds and 2 random seeds,  under a smaller sample size (n), lower dimensionality (p) and higher estimated noise ($\hat{\sigma}_{\e}$)}
\centering
\fontsize{10pt}{10pt}\selectfont
\begin{tabular}{ l l r  r  r  c  c  c  c  c  c  c }
\toprule
Dataset & Source & n & p & $\hat{\sigma}_{\e}$ & Missing & TRUST & RF & M5' & CART & Lasso & NH \\ \midrule
Bones & Hastie et al. \cite{Hastie2009bis} & 485 & 2 & 99.1 & 0 & \textbf{0.66} & \textbf{0.66} & 0.79 & 0.67 & 0.76 & \textbf{0.66} \\ 
Bones2 & Hastie et al. \cite{Hastie2009bis} & 485 & 2 & 100.4 & 225 & \textbf{0.70} & 0.73 & 0.79 & 0.82 & 0.81 & 0.75 \\ 
CorrelatedN & Synthetic & 500 & 8 & 259.6 & 0 & 0.75 & \textbf{0.67} & 0.81 & 0.78 & 0.95 & 0.74 \\ 
MaxN & Synthetic & 500 & 4 & 83.3 & 0 & 0.51 & 0.51 & 0.69 & 0.57 & \textbf{0.50} & 0.57 \\ 
Servo & UCI repository & 167 & 4 & 51.0 & 0 & \textbf{0.15} & 0.32 & 0.66 & 0.36 & 0.64 & 0.38 \\ 
StepsN & Synthetic & 500 & 4 & 110.7 & 0 & \textbf{0.44} & 0.47 & 0.59 & 0.49 & \textbf{0.44} & 0.54 \\ 
\midrule
Mean & & 440 & 4 & 117.4 & 0 & \textbf{0.54} & 0.56 & 0.72 & 0.62 & 0.68 & 0.61 \\ 
Std & & 55 & 1 & 29.7 & 0 & 0.09 & 0.06 & 0.04 & 0.07 & 0.08 & 0.06 \\ 
Mean Rank & &  &  &  &  & \textbf{1.50} & 1.83 & 5.50 & 4.17 & 3.83 & 3.17 \\ 
Std Rank & &  &  &  &  & 0.34 & 0.31 & 0.34 & 0.40 & 0.91 & 0.60 \\ 
\bottomrule
\end{tabular}
\label{tab:resultsG7}
\end{table}

\begin{table}[H]
\caption{Mean unexplained variance by model across 10 folds and 2 random seeds,  under a smaller sample size (n), lower dimensionality (p) and lower estimated noise ($\hat{\sigma}_{\e}$)}
\centering
\fontsize{10pt}{10pt}\selectfont
\begin{tabular}{ l l r  r  r  c  c  c  c  c  c  c }
\toprule
Dataset & Source & n & p & $\hat{\sigma}_{\e}$ & Missing & TRUST & RF & M5' & CART & Lasso & NH \\ \midrule
Correlated & Synthetic & 500 & 8 & 42.7 & 0 & 0.37 & \textbf{0.34} & 0.51 & 0.52 & 0.94 & 0.60 \\ 
Fuel & R package alr4 & 51 & 6 & 7.9 & 0 & \textbf{0.08} & 0.14 & 0.08 & 0.52 & \textbf{0.08} & \textit{N/A} \\ 
Galaxy & UCI repository & 323 & 4 & 0.7 & 0 & \textbf{0.03} & \textbf{0.03} & >1 & 0.10 & 0.11 & 0.07 \\ 
Inflation & European Central Bank & 326 & 7 & 0.2 & 0 & \textbf{0.05} & >1 & \textbf{0.05} & >1 & 0.13 & >1 \\ 
Land & R package alr4 & 67 & 4 & 20.7 & 0 & \textbf{0.23} & 0.37 & >1 & 0.42 & 0.25 & 0.33 \\ 
Lungs & R package isdals & 654 & 4 & 14.5 & 0 & \textbf{0.23} & 0.29 & >1 & 0.27 & 0.24 & 0.25 \\ 
Max & Synthetic & 500 & 4 & 17.9 & 0 & \textbf{0.04} & 0.13 & 0.52 & 0.19 & 0.12 & 0.27 \\ 
Mpg & UCI repository & 398 & 7 & 10.1 & 6 & 0.17 & \textbf{0.13} & >1 & 0.23 & 0.20 & 0.19 \\ 
Sleep & R package ggplot2 & 83 & 7 & 0.1 & 136 & \textbf{0.00} & 0.11 & \textbf{0.00} & 0.08 & \textbf{0.00} & 0.05 \\ 
Steps & Synthetic & 500 & 4 & 59.3 & 0 & \textbf{0.18} & 0.22 & 0.45 & 0.27 & \textbf{0.18} & 0.36 \\ 
\midrule
Mean & & 340 & 6 & 17.4 & 14 & \textbf{0.14} & 0.28 & 0.56 & 0.36 & 0.23 & 0.35 \\ 
Std & & 67 & 1 & 6.2 & 14 & 0.04 & 0.09 & 0.13 & 0.09 & 0.08 & 0.10 \\ 
Mean Rank & &  &  &  &  & \textbf{1.20} & 3.20 & 4.20 & 4.40 & 2.70 & 3.89 \\ 
Std Rank & &  &  &  &  & 0.13 & 0.55 & 0.76 & 0.16 & 0.56 & 0.31 \\ 
\bottomrule
\end{tabular}
\label{tab:resultsG8}
\end{table}

\section{Conclusion}
\label{sec:Conclusion}

This work introduced TRUST (Transparent, Robust, and Ultra-Sparse Trees), a novel regression tree algorithm that combines interpretability, robustness, and predictive accuracy. TRUST is inherently interpretable due to its axis-aligned splits and additive piecewise-linear structure. Transparency is further enhanced through (i) a new variable importance scoring method that accounts for feature correlations (unlike permutation-based methods), includes uncertainty quantification, and avoids the biases common in the majority of other methods; and (ii) the integration of Gemini (Google’s flagship LLM) to generate custom, user-friendly explanations for individual predictions.

TRUST features enhanced robustness via real-time out-of-distribution detection, extrapolation error mitigation through dynamic leaf-level prediction truncation, organic handling of missing values -- including pattern learning from missingness without added computational cost -- and efficient performance in high-dimensional settings. At each leaf, relaxed Lasso models are fit to ensure maximal sparsity without sacrificing accuracy.

Extensive validation on both synthetic and real-world benchmarks shows that TRUST consistently outperforms other interpretable models in predictive accuracy, while rivaling that of Random Forest  and substantially improving upon conceptually related models such as M5'. Given this combination of accuracy and interpretability, TRUST is especially well suited for high-stakes domains like healthcare, finance, and public policy. 

Future work includes extending the framework to classification tasks.

\bibliographystyle{splncs04}
\bibliography{references}

\end{document}